\documentstyle[12pt]{article}
\title{\nopagebreak
\begin{flushright}
\tenrm UCTP116.99
\end{flushright}\vskip0.3in
\nopagebreak
\large \bf Spinning Charged  Solutions in
2+1 Dimensional Einstein-Maxwell-Dilaton Gravity}
\author{Sharmanthie Fernando\thanks{email address:
fernando@physung.phy.uc.edu}\\
\it \small \it Physics Department, University of Cincinnati,
Cincinnati, OH 45221, USA}
\date{}

\begin{document}
\maketitle

\begin{abstract}
We report a new class of rotating charged 
solutions in 2+1 dimensions. 
These  solutions are obtained for   
Einstein-Maxwell 
gravity coupled 
to a dilaton field with 
selfdual electromagnetic fields. 
The mass and the angular momentum of these solutions
computed at
spatial infinity are finite.
The class of solutions considered here have 
naked singularities
and are asymptotically flat.
\end{abstract}

\section{Introduction}

Interest in 2+1 dimensional gravity has been heighten
by
the discovery of a black hole solution by Ba\~{n}ados {\it et al.}
\cite{banados1}.
This black hole, (named BTZ) has anti-de Sitter structure
locally and globally differ to anti-de Sitter by identifications
done with a discrete subgroup of the isometry group
of anti-de Sitter space, $SO(2,2)$ \cite{banados2}. It enjoys many
black hole properties of its counterparts in higher dimensions
which makes BTZ a suitable model to understand black hole
physics in a technically simpler setting.

Extension of the BTZ black hole with charge have been met with mixed success. 
The first  investigation into 
static  charged black holes  was done by Ba\~{n}ados {\it et al.}\cite{banados1}.
Due to the 
logarithmic  nature of the electromagnetic  potential, these solutions gave 
rise to 
unphysical properties \cite{hirsh}. The horizonless static  solution with  
magnetic charge 
were studied by Hirshmann {\it et al.}\cite{hirsh}, and the persistence  of 
these unphysical 
properties  was highlighted by Chan\cite{chan1}. Kamata {\it et al.}\cite{kam1}
 presented a  rotating 
charged black hole with self (anti-self) duality imposed on  the  electromagnetic fields.
 The resulting solutions  were asymptotic to an extreme BTZ black hole solution 
 but had  diverging mass and angular momentum \cite{chan1}. 
 Cl\'ement\cite{clem}, Fernando and Mansouri\cite{fer},
introduced a Chern-Simons term as a regulator to screen the electromagnetic
 potential
 and obtained horizonless
 charged solutions.

In this work, we couple a dilaton
to  Einstein-Maxwell gravity to obtain
rotating charged solutions with finite
mass and finite angular momentum.
It is well known that the introduction of 
a dilaton field drastically changes the 
space-time structure in
3+1 dimensions. Furthermore, studying  dilaton gravity is important
since it arises in low energy string theory.
Therefore it is worthwhile to see how the 
presence of the dilaton would
modify the BTZ black hole solutions and how
it would help in curing the divergences
occurred in the previous charged
solutions.

Einstein-Maxwell-Dilaton action in 2+1 dimensions is written as follows:
\begin{equation}
I = \int d^3x \sqrt{|g|}\left[R - 2\Lambda e^{\beta\phi} - 
\frac{\gamma}{2} (\bigtriangledown^{\mu} \phi \bigtriangledown_{\mu}\phi ) - e^{-4\alpha\phi}F^2\right] 
\end{equation}
Here $\Lambda$ is the cosmological constant and  we consider
the case for $\Lambda < 0$ which 
corresponds to anti-de Sitter spaces.
$\beta, \gamma, \alpha$ are coupling constant.
In this expression $1/16 \pi G$ is taken to be 1.
The above action is the low energy string action when $\gamma=8$, $\beta=4$, 
$\alpha=1$. There are several work related to 
charged solutions to the above action.
Chan {\it et al.}\cite{chan2} have studied a one parameter 
family of static charged dilaton 
black holes which were non-asymptotically anti-de sitter and solutions 
with cosmological horizons
when $4\alpha=\beta$.
Park {\it et. al.}\cite{park} obtained axially symmetric static solutions by using a 
dimensional reduction method. 
A more general magnetically charged solution to dilaton gravity
 was found by Koikawa {\it et al.}\cite{kam2}.
 By applying T-duality to the static electric charged black holes of
 Chan {\it et al.}\cite{chan3}, Chen\cite{chen} obtained  rotating charged black hole solutions
 to Einstein-Maxwell-Dilaton gravity. However, these solutions
 are not the most general rotating charged solutions to 
 Einstein-Maxwell-Dilaton gravity.

The aim of the present paper
is to search for rotating charged solutions
to Einstein-Maxwell-Dilaton gravity with self duality
imposed on the electromagnetic  fields. 
The plan of the paper is follows. In section 2 we will
compute the most general solutions with self duality
imposed. In section 3 we will impose restrictions on coupling constants
so that the mass and angular momentum are finite. In section 4 we will
discuss the properties of theses solutions and finally
give concluding remarks.

\section{General Solutions}

By extremizing the Lagrangian of the action in equation(1) 
with respect to the metric $g_{\mu \nu}$, the electromagnetic potential 
$A_{\mu}$ and the dilaton field $\phi$,
one obtains the corresponding field equations for gravitational, electromagnetic and 
dilaton respectively 
as follows.
\begin{equation}
R_{\mu\nu}= - 2\Lambda g_{\mu\nu}e^{\beta\phi} + 
e^{-4\alpha \phi}(2F_{\mu\rho}F_{\nu}^{\rho} - 
g_{\mu\nu}F_{\lambda\sigma}F^{\lambda\sigma}) + \frac{\gamma}{2} (\bigtriangledown_{\mu} \phi)(\bigtriangledown_{\nu} \phi) 
\end{equation}
\begin{equation}\bigtriangledown_{\mu}(e^{-4\alpha \phi}F^{\mu\nu}) =0
\end{equation}
\begin{equation}
\frac{\gamma}{2} \bigtriangledown^{\mu}\bigtriangledown_{\mu} \phi + 
2\alpha e^{-4\alpha\phi} F^2 - \beta e^{\beta \phi} \Lambda = 0
\end{equation}
Let us assume that the three dimensional space-time is stationary and circularly  symmetric, having  two commuting Killing vectors, $\frac{\partial}{\partial \phi}$ and $\frac{\partial}{\partial t}.$ Such a space-time could be parameterized with a line element as follows:
\begin{equation}
ds^2=  - N^2dt^2  + L^{-2}d\rho^2 + K^2(d\theta + N^\theta dt)^2
\end{equation}
The functions $N,L,K$ and $N^\theta$  depends only  on radial coordinate $\rho$.
We use the tetrad formalism and Cartan structure equations to look for 
solutions. Obvious non coordinate basis for the above metric would be,
\begin{equation}
e^0  = Ndt ;\hspace{1.0cm}  e^1 = K (d\theta + N^\theta dt) ; \hspace{1.0cm}  
e^2= L^{-1}d\rho
\end{equation}
The indices $a,b = 0,1,2$ are for the orthonormal basis and $\mu, \nu = 0,1,2$
for the coordinate basis with $x^0 = t$; 
$x^1 = \theta$ ; $x^2 = \rho$. The non-vanishing components of the 
electromagnetic field tensor in the coordinate basis are given by
$F_{t\rho} = E$, $F_{\rho \theta} = B$,
and in the orthonormal basis they are given by $F_{02} = \hat{E}$,
$F_{21} = \hat{B}$.
They are related by,
\begin{equation}
E= \frac {(\hat{E}N - \hat{B}KN^{\theta})} {L};
\hspace{1.0cm}
B = \frac {\hat{B}K} {L}
\end{equation}
From  electromagnetic field equations(3), 
\begin{equation}
F^{t\rho} = \frac{C_1 e^{4\alpha\phi}} {\sqrt{|g|}};
\hspace{1.0cm}
F^{r \theta} = \frac{C_2 e^{4\alpha\phi}} {\sqrt{|g|}}
\end{equation}
In order to seek solutions to the 
gravitational field equations, we make the
ansatz that, the electric and the magnetic fields are self dual in 
the orthonormal
basis. Therefore,
\begin{equation}
\hat{E}  = \hat{B}= u(\rho)
\end{equation}
With the use of equations(7,8,9),
$$u(\rho) =  -C_1 \frac {\exp[4\alpha \phi]} {K}$$
\begin{equation}
N^{\theta} =  \frac{N}{K} -\frac{C_2}{C_1}
\end{equation}
Considering the behavior of $u=\hat{E}$
for  flat space-times, we can represent $C_1$ with
electric charge $Q_e$.
A value for $C_2$ will be assigned later.

Having used the ansatz of self duality, the gravitational field equations 
in the orthonormal basis 
takes the following form for a circularly symmetric space time.
\begin{equation}
R_{00} = L^{2} \left(\frac{N''}{N} + \frac{N'K'}{NK} \right) + 
\frac{LL'N'} {N} - 2\left(\frac{KLN^{\theta'}}{2N}\right)^2  = 
- 2\Lambda e^{\beta\phi} + 2e^{-4\alpha \phi}u^2 
\end{equation}
\begin{equation}
R_{11} =  -L^{2} \left(\frac{K''}{K} + \frac{N'K'} {NK} \right) - 
\frac{LL'K'} {K} - 2\left(\frac{KLN^{\theta'}}{2N}\right)^2  = 
2\Lambda e^{\beta\phi} +  2e^{-4a\phi}u^2 
\end{equation}
\begin{equation}
R _{01} = -\frac{L}{K^2} \left(\frac{K^3 LN^{\theta'}}{2N}\right)'  = 
-2e^{-4\alpha\phi}u^2
\end{equation}
\begin{equation}
G_{22} =  L^{2} \left( \frac{N'K'} {NK} \right) + 
\left(\frac{KLN^{\theta'}}{2N}\right)^2  =  -\Lambda e^{\beta\phi}  + 
\frac{\gamma}{4} (\hat{\bigtriangledown}_{2}\phi)^2 
\end{equation}
The gravitational field  equations(11-14), the dilaton
field equation(4) and the conditions (10)  
leads to the final equations to be solved as follows:
\begin{equation}
(LX')' = -4\Lambda e^{\beta \phi} \frac{X}{L}
\end{equation}
\begin{equation}
L( H K^2)' = 2Q_e^2 e^{4\alpha \phi}
\end{equation}
\begin{equation}
\left(\frac{X'L}{2X}\right)^2 = \frac{\gamma}{4} (\phi' L )^2 - 
\Lambda e^{\beta \phi}
\end{equation}
\begin{equation}
\gamma( \phi LX )' = 2\Lambda \beta e^{\beta \phi} \frac{X}{L}
\end{equation}
where,
\begin{equation}
X =NK;     \hspace{1.0cm}  
H = \frac{L}{2} \left(\frac{N'}{N}  - \frac{K'}{K}\right)
\end{equation}
Here, the prime is given by $d/d\rho$. Equation(15) is obtained
by observing that $R_{00} + 2\Lambda e^{\beta \phi} =
R_{11} - 2\Lambda e^{\beta \phi}$. Equation(16) is obtained by
the fact that $R_{00} + R_{11} = -2R_{01}$.
Equation(17) is just equation(14) rewritten.
Equation(18) is the dilaton field equations with the self duality imposed.
The  solution to the above set of equations are,
$$
X = a_1 exp( \frac{-2\gamma \phi}{\beta})
$$
$$
L = \frac{exp(\frac{\beta \phi}{2})}{b_0 \phi'}
$$
\begin{equation}
K^2 =  a_2 exp{(\frac{-2\gamma \phi}{\beta})} - 
\frac{2 b_0 a_3}{ (\frac{2\gamma}{\beta} 
- \frac{\beta}{2})} exp(\frac{-\beta \phi}{2})
-\frac{4Q_e^2 b_0^2}{(4\alpha -\frac{\beta}{2}) 
(4\alpha - \beta + \frac{2\gamma}{\beta})} 
exp((4\alpha - \beta) \phi)
\end{equation}
Here $b_0^2 = \frac{(4\gamma^2 - \gamma \beta^2)}
{(4|\Lambda|\beta^2)}$
and the above solutions are valid only when $b_0^2 > 0$,
$(4\alpha-\frac{\beta}{2}) \neq 0$,
$(4\alpha -\beta + \frac{2\gamma}{\beta}) \not= 0$,  and 
$\beta \not= 0$.
Here $a_1, a_2, a_3$ are integrating constants.

From the above equations it is clear that the all the functions 
$N,L,K,N^{\theta}$ depend
on the dilaton field $\phi$. In order to compare these solutions with the 
previously obtained
dilaton solutions and also to see the correspondence with the BTZ 
black hole we pick  
$L$ to be the following.
\begin{equation}
L = \frac{|\Lambda|^{1/2}(\rho^2 - \rho_0^2)^{\omega}} {\rho}
\end{equation}
Here, $\omega$ and $\rho_0^2$
are  constants which would be determined by imposing some
restrictions on the solutions to give finite mass and 
finite angular momentum.
Note that equation(21) is equivalent to  the dilaton field $\phi$  and
$X$ been,
\begin{equation}
\phi = ln \left(b_1 (\rho^2 -\rho_0^2)^{\frac{2(\omega-1)}{\beta}}\right);
\hspace{1.0cm} X = |\Lambda|^{1/2} ( \rho^2 -\rho_0^2)^{\frac{4 \gamma(1-\omega)}{\beta^2}}
\end{equation}
where
$b_1 = [\frac{-\beta}{4b_0(-\omega + 1) |\Lambda|^{1/2}}]^{-2/\beta}$
and $a_1$ is normalized so that $a_1 b_1^{\frac{-2\gamma}{\beta}} = 
|\Lambda|^{1/2}.$
By substituting  $\phi$ into equation(20),
\begin{equation}
K^2 =  \bar{a}_2 (\rho^2 - \rho_0^2)^{\frac{4\gamma(-\omega+1)}{\beta^2}} 
+ \overline{a}_3(\rho^2-\rho_0^2)^{(-\omega +1)}
+ \overline{Q}_e^2(\rho^2 -\rho^2_0)^{\frac{2(4\alpha - \beta)(\omega - 1)}{\beta}}
\end{equation}
where $ \overline{a}_2 = a_2 b_1^{-2\gamma/\beta}$,
$\overline{a}_3 = \frac{-2 b_0 b_1^{(-\beta/2)}a_3}
{ (\frac{2\gamma}{\beta} 
- \frac{\beta}{2})}$,
$\overline{Q}_e^2= \frac{-4Q_e^2 b_0^2 b_1^{4\alpha - \beta}}
{(4\alpha -\frac{\beta}{2})
 (4\alpha - \beta + \frac{2\gamma}{\beta})}$.
We restrict  $\omega \not=1$. Otherwise the dilaton would be
a constant and the solutions would be trivial.

\section{Quasilocal Mass and Angular Momentum}

In this section, we will impose
restrictions on the values
of $\alpha, \beta, \gamma, \omega$
so that the above solutions
have finite mass and
finite angular momentum.
Also we will interpret the integrating constants
$a_1, a_2, a_3$ as appropriate
physical constants.
Here we will adopt the prescription of Brown {\it et.al.}
\cite{bro1} \cite{bro2} in computing mass and angular momentum.

\subsection{Quasilocal Angular Momentum}
The quasilocal angular momentum according to the prescription
given by Brown {\it et.al.}\cite{bro1} \cite{bro2}
is,
\begin{equation}
J(\rho) = \frac{K^3 N^{\theta '}L}{N}
\end{equation}
For the metric in equation(20) it is equivalent to,
\begin{equation}
J(\rho)= -2 H K^2
\end{equation}
From equations(19) and (23),
\begin{equation}
K^2H = -|\Lambda|^{1/2}\overline{a}_3 -
2|\Lambda|^{\frac{1}{2}} \overline{Q}_e^2 ( \frac{4\alpha}{\beta} - 1)(\omega - 1)
(\rho^2 -\rho^2_0)^{\frac{(8\alpha - \beta)(\omega - 1)}{\beta}}
\end{equation}
Hence for large $\rho$, $J(\rho)$ 
becomes  finite only if  
\begin{equation}
(\omega - 1 ) ( \frac{8\alpha}{\beta} - 1) \leq 0
\end{equation}
Since $\omega \not= 1$ and $(8\alpha - \beta) \not= 0$,
the present class of solutions will obey strictly smaller
condition for equation(27).
Hence,
\begin{equation}
\lim_{\rho \rightarrow \infty} J(\rho) = J  = 2|\Lambda|^{\frac{1}{2}} \overline{a}_3
\end{equation}

\subsection{Quasilocal Mass}

From the definition of Brown {\it et. al.} \cite{bro1}\cite{bro2}, 
the quasilocal mass is given by,
\begin{equation}
M(\rho) = 2 N(\rho) \left[L_0(\rho) - L(\rho)] \right) - 
J(\rho) N^{\theta}(\rho)
\end{equation}
Here, $L_0(\rho)$ is chosen to be the reference when there is zero mass. 
In comparison with the BTZ black hole  we choose 
\begin{equation}
L_0 = (L)_{\rho_0 = 0} = |\Lambda|^{1/2}\frac{\rho^{2 \omega}}{\rho}
\end{equation}
Consider the first term in equation(29),
\begin{equation}
2N ( L_0 - L)= \frac{2|\Lambda|^{\frac{1}{2}}X}{K} \left( \omega \rho_0^{2}
\rho^{(2 \omega - 3)} + \mbox{lower order terms of $\rho$ } \right)
\end{equation}
For finite mass, $(X/K)$ should behave as $\rho^n$ with  $n \leq 
-(2 \omega -3)$.
If $n < - (2 \omega -3 )$, then this
solution will correspond to the ``massless'' BTZ solution for appropriate 
limits.
Hence to avoid such extreme cases, we would pick $n$
to be $ ( -2\omega + 3)$.
Therefore, since at large $\rho$, 
\begin{equation}
X \rightarrow \rho^{\frac{8\gamma(-\omega+1)}{\beta^2}}
\end{equation} 
$K$ should behave as,
\begin{equation}
K \rightarrow  \rho^{(\frac{8\gamma(-\omega + 1)}{\beta^2} +(2\omega -3))}
\end{equation}
for large $\rho$ for $n = (-2\omega + 3)$.
Now, we will consider $N^{\theta}$ term at large $\rho$.
\begin{equation}
N^{\theta} =  \frac{X}{K^2} - \frac{C_2}{C_1}
\end{equation}
To impose the boundary condition $N^{\theta}(\infty) =0$,
$X/K^2$ has to converge for $\rho \rightarrow \infty$.
With the constraint in equation(33) it means,
\begin{equation}
(-\omega + 1 ) ( -\frac{8\gamma}{\beta^2} + 4) + 2 \leq 0
\end{equation}
To approximate the behavior of $K^2$ to be $\rho^2$ at large
and small $\rho$, we assume all terms in $K^2$ has positive powers
of $\rho^2$.
Considering the fact that $\overline{a}_3 (\rho^2-\rho_0^2)^{(-\omega +1)}$ 
term in $K^2$ is proportional to the
angular momentum,
to take appropriate limits, 
we take $\bar{a}_2 (\rho^2 - \rho_0^2)^{\frac{4\gamma(-\omega+1)}{\beta^2}}$
term to be the dominant power of $K^2$.
Without loss of generality we assume $\overline{a}_2 =1$.
Hence
$$
\frac{4\gamma(-\omega +1)}{\beta^2} > (-\omega +1) >0
$$
\begin{equation}
\frac{2(4\alpha - \beta)(\omega - 1)}{\beta}>0
\end{equation}
and from equation(33) and assumptions in equation(36),
$\omega = 
\frac{(4 \gamma - 3\beta^2)}{(4 \gamma - 2 \beta^2)}$.
With all these preliminaries,
\begin{equation}
\lim_{\rho \rightarrow \infty} N^{\theta}(\rho) = (|\Lambda|^{1/2} - C_2/C_1)
\end{equation}
To impose $N^{\theta}(\infty)$ = 0,
we let $C_2 = |\Lambda|^{1/2}C_1$. Hence,
\begin{equation}
\lim_{\rho \rightarrow \infty}  M(\rho) = M = 
2 |\Lambda| \omega \rho_0^2 
\end{equation}
leading to $\rho_0^2 = \frac{M} {2 |\Lambda| \omega}$.

\section{Exact Solutions with Finite Mass and Finite Angular momentum}

With the above preliminaries, the final form of 
the solutions with finite mass and finite angular momentum
is,
$$
L = \frac{|\Lambda|^{1/2}(\rho^2 - \rho_0^2)^{\omega}} {\rho}
$$
$$
N = \frac{|\Lambda|^{1/2}(\rho^2 - \rho_0^2)^{(-2\omega+3)}}{K}
$$
$$
N^{\theta} = \frac{N}{K} - |\Lambda|^{1/2}
$$
\begin{equation}
K^2 =  (\rho^2 - \rho_0^2)^{(-2\omega + 3)} 
+ \overline{a}_3 (\rho^2-\rho_0^2)^{(-\omega + 1)}
+\overline{Q}_e^2 
(\rho^2 -\rho^2_0)^{\frac{2(4\alpha - \beta)(\omega -1)}{\beta}}
\end{equation}
\begin{equation}
\phi = ln \left(b_1 (\rho^2 -\rho_0^2)
^{\frac{2(\omega -1)}{\beta}}\right)
\end{equation}
With $\omega = \frac{(4 \gamma - 3 \beta^2)}{(4 \gamma - 2 \beta^2)}$,
$\rho^2_0 = \frac{M}{2 |\Lambda| \omega}$ and
$\overline{a}_3 = \frac{J}{2 |\Lambda|^{\frac{1}{2}}}$.
The electromagnetic potential is
\begin{equation}
A = A_{\mu} dx^{\mu} = \frac{(\rho^2 -\rho_0^2)^m}{2m} \left( |\Lambda|^{1/2} dt + d\theta \right).
\end{equation}
where $m = \frac{\beta(8\alpha -\beta)}
{2(\beta^2 - 2 \gamma)}$.
From the constraint imposed in equation(27),
it is obvious that $m < 0$. Hence, the potential is finite
at large $\rho$. Therefore, the presence of the dilaton
``screens'' the electromagnetic potential.
To clarify this further, if we  look at dilaton
field in flat space for self-dual case as considered above,
the values for $\hat{E}$ and  $\hat{B}$ would be  
$Q_eexp[4\alpha b_0]/\rho$.
Hence the potential $A_t = Q_er^{4\alpha b_0}/(4\alpha b_0)$
is finite for $4\alpha b_0 < 0$. Therefore
the
presence of the dilaton modifies the Coulomb force in 2+1 dimensions.
We may recall that
in \cite{clem}\cite{fer},
a topological mass term $m_p \epsilon^{\alpha \beta \gamma}
A_{\alpha} F_{\beta \gamma}$
was introduced to Einstein-Maxwell gravity
to cure the divergences of the quasilocal mass $M$ which gives a similar effect
to the Coulomb force.
If we compare the
electric field  in three separate cases,
\begin{equation}
E_{Coulomb} = \frac{Q_e}{\rho} \hspace{1.0cm} 
E_{Dilaton} = \frac{Q_e \rho^{4\alpha b_0}}{\rho}
\hspace{1.0cm} E_{topological} = \frac{Q_e e^{-m_p\rho}}{\rho}
\end{equation}
the topological mass term has a better regulating effect
in comparison with the dilaton field.

\subsection{Causal Structure}

The curvature scalars in 2+1 dimensions are $R$, 
$R_{\alpha \beta}R^{\alpha \beta}$
and $det R_{\alpha \beta}/det g_{\alpha \beta}$.
For the above solutions,
with $\gamma, \beta, \alpha \neq 0$ and
for $M > 0$, 
all of them diverge at $\rho = \rho_0$ and finite everywhere else.
Hence the curvature 
singularity at $\rho = \rho_0$ is a naked singularity
without horizons. However,
$K^2$ which is the $g_{\theta \theta}$
term in the metric 
has to be positive to avoid closed time like
curves since $\theta$ is a periodic coordinate.
Even with the constraints
imposed on the parameters $\alpha$, $\beta$, $\gamma$, 
there is a possibility that $K^2$
would become negative. Hence
one has to include the possibility of closed
time like curves for these solutions.
The scalar curvature $R$ is $(6\Lambda + \frac{\gamma}{4 b_0^2})
(\rho^2 -\rho_0^2)^{\frac{(\omega-1)}{2}}$.
Since $(1-\omega)> 0$, $R \rightarrow   0$ for large $\rho$.
Therefore the solution  becomes flat asymptotically.

\subsection{The relation with  BTZ Black Hole}
Note that above discussion is for when
$\gamma$, $\beta$ and $\alpha$ is non vanishing.
To see the correspondence of the 
above solutions with BTZ black hole,
let us take the limit $\omega \rightarrow 1$ and  $Q_e \rightarrow 0$.
Then,
\begin{equation}
M \rightarrow 2|\Lambda| \rho_0^2;
\hspace{1.0cm} \phi \rightarrow 0 ;
\hspace{1.0cm}
K^2 \rightarrow (\rho^2 - \rho^2_{0}) + \overline{a}_3
\end{equation}
If $\rho_0^2 = \overline{a}_3$ then
$M= |\Lambda|^{1/2}J$ with the following metric,
$$L = N \rightarrow \frac{|\Lambda|^{1/2}(\rho^2 - \frac{M}{2 |\Lambda|^{1/2}})} {\rho}$$
$$ K\rightarrow \rho$$
\begin{equation}
N^{\theta} \rightarrow \frac{J}{2\rho^2}
\end{equation}
Hence the solution obtained in this work
approaches an
extreme BTZ black hole as a special case. The presence
of the dilaton and charge have left
the BTZ space-time  horizonless and asymptotically flat.

\section{Conclusions}

We have obtained a family
of rotating
charged solutions
to   Einstein-Maxwell-Dilaton gravity in 2+1 dimensions.
Here we have imposed self duality
on the electric and magnetic field to facilitate
solve the field equations exactly. With certain constraints on the coupling
constants $\gamma$, $\beta$ and $\alpha$,
we obtained solutions with finite mass and finite
angular
momentum. For non-zero values of $\alpha$ $\beta$ and $\gamma$,
the class of solutions considered in this paper
are horizonless, have naked
singularities and  are asymptotically flat.
These solutions approaches
the extreme BTZ black hole solutions
as a special case for $\alpha, \beta, \gamma \rightarrow 0$.
The
presence of the dilaton ``screens'' the electromagnetic potential
and modifies the structure of the space-time considerably.
However, since the metric depends on the dilaton as it is clear
from equation(20), one may use other polynomial functions
for the  dilaton
field in these solutions to understand how the space-time structure changes
accordingly. It is also a question   how a massive dilaton would effect
the space-time structure of  the above solutions.
In extending this work it may be possible
to include a potential
of the form $V(\phi)= 2\Lambda_1e^{\beta_1 \phi} + 2 \Lambda_2 e^{\beta_2 \phi}$
to the action considered in this paper. These kind of potentials are investigated in 
dimensions $n \geq 4$ and have shown the possibility of having three
horizons by Chan {\it et al.} \cite{chan3}.
It would be interesting to see whether
one can construct  rotating charged dilaton
black holes with the above
potential in 2+1 dimensions.

We may recall
that Chen \cite{chen} obtained
rotating charged black hole solutions for
dilaton gravity with $\gamma$ = 4 and $ 4\alpha = \beta$
by a T-duality transformation on the static charged
black holes of Chan {\it et al.} \cite{chan2}.
The solutions obtained in this paper are for more
general values of 
$\alpha$, $\beta$, $\gamma$ and 
they are not T-dual to the solutions of Chen \cite{chen}.
Furthermore instead of Maxwell fields,
one can include Yang-Mills fields
to study the space-time structure for dilaton
gravity in  2+1 dimensions which we hope to report elsewhere.

{\bf Acknowledgments}: I wish to thank F. Mansouri
and M. Muhkerjee for helpful comments. This work was supported in part
by the Department of Energy under contract number DOE-FG02-84ER40153.

\end{document}